\documentclass[11pt]{iopart}

\usepackage{graphicx}
\usepackage{dcolumn}
\usepackage{bm}
\usepackage{citesort}
\begin{document}

\title[Magnetic phase transitions in GdSc using non-contact utrasonics]{Magnetic phase transitions in Gd$_{64}$Sc$_{36}$ studied using non-contact ultrasonics}

\author{Oksana Trushkevych, Yichao Fan, Robert Perry and Rachel S. Edwards}
\address{Department of Physics, University of Warwick, Coventry, CV4 7AL, UK}
\ead{o.trushkevych@warwick.ac.uk}

\begin{abstract}
The speed and attenuation of ultrasound propagation can be used to determine material properties and identify phase transitions.  Standard ultrasonic contact techniques are not always convenient due to the necessity of using couplant, however, recently reliable non-contact ultrasonic techniques involving electromagnetic generation and detection of ultrasound with electromagnetic acoustic transducers (EMATs) have been developed for use on electrically conducting and/or magnetic materials.  We present a detailed study of magnetic phase transitions in a single crystal sample of Gd$_{64}$Sc$_{36}$  magnetic alloy using contact and non-contact ultrasonic techniques for two orientations of external magnetic field.  Phase diagrams are constructed based on measurements of elastic constant $C_{33}$, the attenuation, and the efficiency of generation when using an EMAT. The EMATs are shown to provide additional information related to the magnetic phase transitions in the studied sample, and results identify a conical helix phase in Gd$_{64}$Sc$_{36}$  in the magnetic field orientation $\vec{H} \| c$.

\end{abstract}

\maketitle

\section{Introduction}

Magnetic alloys based on rare earth metals  have attracted much attention, due to having a variety of magnetic phases \cite{nature07}, and to their potential applications in magnetic refrigeration and thermomagnetic generation \cite{Elliott59,Hsu11,Sandeman12}. Gd and Sc alloys with compositions around Gd$_{70}$Sc$_{30}$  exhibit a variety of phases including paramagnetic, ferromagnetic, basal plane helical, and for certain Sc content a conical helical phase \cite{Melville88, Silva95} due to the competition between the ferromagnetic ordering of Gd and the helical antiferromagnetic ordering prevalent in Sc, and to the random distribution of magnetic ions.

\begin {figure} [h]
   \includegraphics[width=70mm]{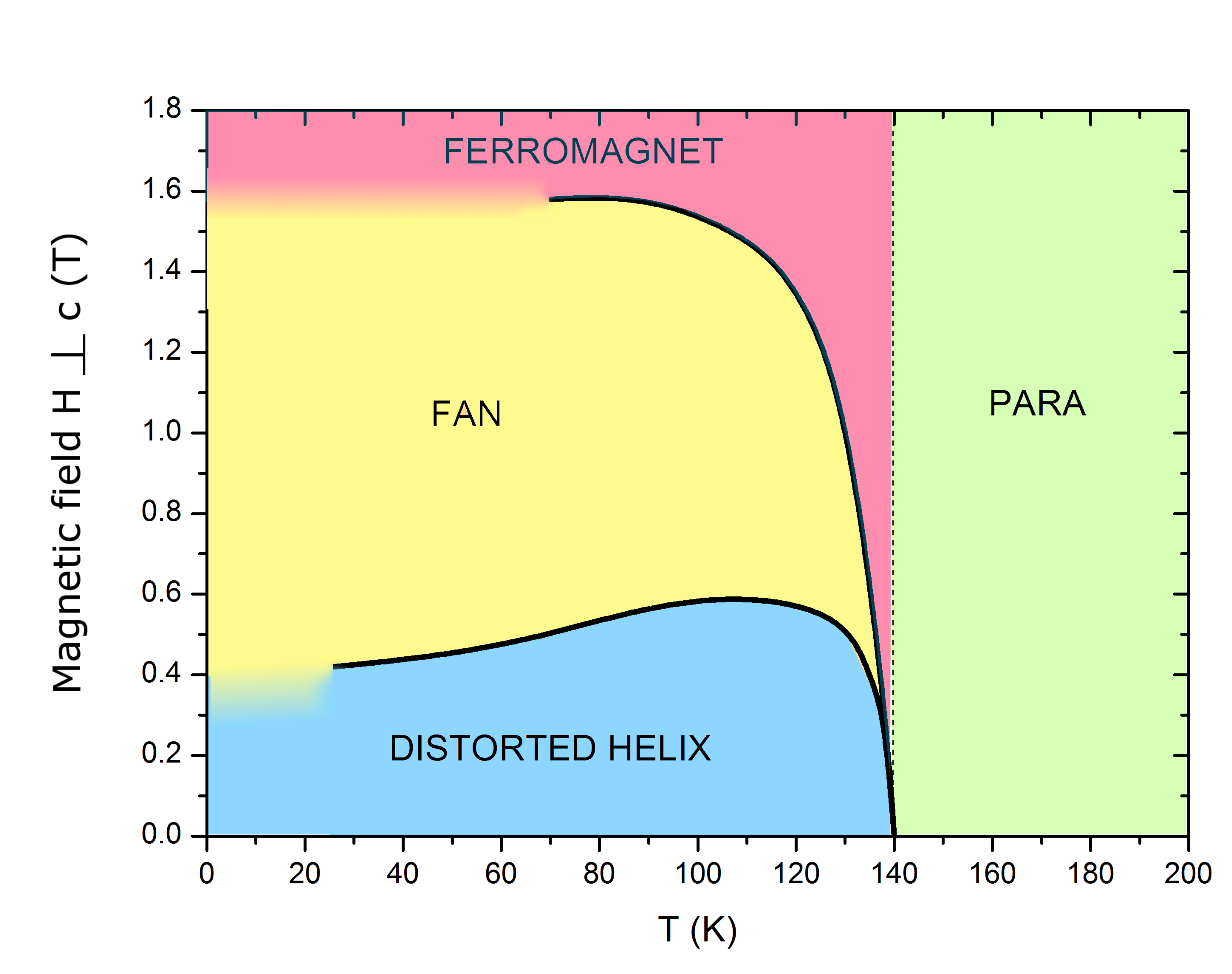}
 \caption {Phase diagram of Gd$_{64}$Sc$_{36}$ crystalline magnetic alloy for $\vec{H}\bot c$, adapted from da Silva et al.\cite{Silva99}
 \label {sketch}}
\end {figure}

The crystalline magnetic alloy  Gd$_{64}$Sc$_{36}$  has a hexagonal close packed (hcp) structure, and was previously studied using neutron diffraction in zero applied magnetic field \cite {Melville88}. This alloy  is paramagnetic above the Neel temperature, $T_N = 140$  K, and at lower temperatures exhibits a basal helical antiferromagnetic (AF) phase with a complex behaviour of the temperature-dependent helimagnetic angle. Magnetic field-induced phases were previously studied for an external magnetic field $\vec{H}$ applied in the $ab$ (basal) plane \cite {Silva99}, with the measured phase diagram shown in Figure \ref{sketch}. At temperatures below $T_N$,   Gd$_{64}$Sc$_{36}$  undergoes a first-order phase transition from helix to fan phase between 0.4 and 0.6~T, and a second order transition to a field-aligned ferromagnetic phase around 1.6~T \cite {Silva99}.

The magnetic phase behaviour  when $\vec{H}$ is appled parallel to $c$ has to now been unknown; the sample exists in an AF helix phase at $T < T_N$ when unperturbed by a magnetic field, and one would expect a field-induced ferromagnetic phase at high magnetic fields. At intermediate fields intermediate phases may exist \cite{Melville88}. Note that, for this field orientation no first-order transition is expected as the magnetic moments for the AF helix arrangement are in the basal plane and perpendicular to the field, and can therefore be continuously rotated to align with the field.

Ultrasonic measurements are known to be a powerful tool for studying structural and magnetic phase transitions \cite{Melville88,Silva95,Silva99,Luthi05b,Kinsler62b}; elastic constants and attenuation values can be extracted from these measurements, and these link to fundamental sample properties. At phase transitions these parameters exhibit anomalies associated with critical phenomena, and transitions are typically accompanied by a softening of certain elastic constants and hence an increase in attenuation. By choosing an appropriate  direction of propagation, wave type (e.g. shear or longitudinal) and polarisation, different elastic constants and anisotropy can be probed \cite{Luthi05b,Kinsler62b}.

Conventional ultrasonic measurements use piezoelectric transducers such as quartz or PZT to generate and detect ultrasound, however, these require the use of couplant and direct physical contact with the sample,\cite{Melville88,Silva95,Silva99,Luthi05b} and with repeated thermal cycling the coupling between the transducer and the sample may quickly deteriorate. Electromagnetic acoustic transducers (EMATs) \cite{Thompson78,Frost79b,Thompson90b}  offer a number of advantages, being non-contact, with narrowband or broadband operation possible through careful transducer design, and have previously been successfully employed to study phenomena such as giant quantum oscillations \cite{Dobbs69} and magnetic phase transitions \cite{Lim98,RachelConf}, as well as being common in non-destructive testing applications \cite{Palmer03}.

In this paper we use both standard (contact quartz) and non-contact (EMAT) transducers to study magnetic phases in  Gd$_{64}$Sc$_{36}$  in a temperature range of 4~-~300~K and for applied magnetic fields of up to 4 T, for two orientations of magnetic field, oriented in the basal plane or parallel to the $c$ axis.  We compare the EMAT efficiency behaviour with previously studied systems \cite{Buchelnikov92,Buchelnikov94,Buchelnikov02}, give a detailed phase diagram for $\vec{H}\| c$, and suggest a correction to the phase diagram for $\vec{H}\bot c$ from reference \cite{Silva99}.

\subsection{EMAT operation}

EMATs consist of a coil of wire and an applied (bias) magnetic field, and their design has been described in detail elsewhere \cite{Palmer03,Dobbs69,RachelConf}. A current is pulsed through the coil, leading to a time-varying magnetic field, and when placed near a sample a mirror current or magnetisation change may occur within the sample; ultrasonic generation is therefore within the sample itself, rather than within the transducer.

In magnetic alloys there are two possible mechanisms for EMAT generation and detection; the Lorentz force and magnetostrictive force mechanisms \cite{Thompson78,Dobbs73b,Jian04,Buchelnikov92}.  In contrast to piezoelectric transducers, which detect either longitudinal or shear waves, the same EMAT can generate and detect shear and longitudinal waves depending on the experimental configuration (coil design and magnetic field direction) and the sample's magnetic state \cite {Dobbs73b}. When generation is by the Lorentz force mechanism, the efficiency scales linearly with the applied magnetic field \cite{Dobbs73b}. However, when generation and detection is solely by magnetostriction, the  efficiency of the EMATs depends on the  magnetisation and magnetostriction constant of the material under investigation.  For a magnetic field-induced  paramagnetic  to ferromagnetic transition, the efficiency usually peaks around the phase transition \cite{Buchelnikov92}.

The efficiency of electromagnetic ultrasonic generation has been studied in various materials, including the rare-earths Tb, \cite{Parkinson77} Gd,  \cite{Andrianov88} and Er. \cite{Buchelnikov02} In these studies the efficiency was found to peak or change abruptly at phase transitions, with a possible explanation for the increase in efficiency as being due to the increased mobility of the spin subsystem close to a magnetic phase transition \cite{Buchelnikov92}. A phase diagram consisting of multidomain, canted and collinear (field aligned ferromagnetic) and paramagnetic phases for pure Gd was previously suggested based solely on the efficiency of EMAT generation \cite{Andrianov88}.  The authors specified two contributions to the EMAT efficiency; one from a paraprocess (orientation in magnetic field of spins that are not aligned in the direction of the resultant magnetization, due to thermal motion) and another from a spin flip process.

Buchel'nikov and Vasiliev \cite{Buchelnikov92} gave an extensive review of EMAT generation of ultrasound and, in particular, discussed the generation of longitudinal waves due to magnetostriction in a monocrystalline ferromagnet with an hcp lattice, in both the paramagnetic and ferromagnetic phases.   Assuming  that the magnetic permeability of the material, $\mu = \mu_0(1+\chi)$,  is large compared to the magnetoelastic interaction parameter $\xi =\frac{\gamma^2 M^2 \chi}{\rho v_l^2}$ (where $\mu_0$ is the magnetic permeability of vacuum, $\chi$ is magnetic susceptibility, $\gamma$ is magnetostriction constant, $M$ is magnetisation, $\rho$ is the mass density, and $v_l$ is the speed of longitudinal sound waves), the efficiency of ultrasonic generation is given by
\begin{equation} \label{result}
\eta \sim \frac{\chi_0\sin^2\Phi \cos^4\Phi}{(1+4\pi \chi_0\cos^2\Phi)^2}, \qquad \chi_0 = \frac{gM}{\omega_2},
\end{equation}
where $A$ is a constant, $\Phi$ is the azimuthal magnetisation angle,  $\chi = \chi_0 \cos^2\Phi$, $\omega_2$ is the frequency of precession of the magnetisation and  describes the antiphase azimuthal oscillations (as expected in the system studied here), $g$ is the magnetomechanical factor and $M$ is the magnetisation \cite{Buchelnikov92}. The dependence on the applied magnetic field is  through $\omega_2 (H)$ and $\Phi(H)$. $\omega_2$ reaches a minimum at the phase transition, hence the susceptibility $\chi_0$ reaches a maximum; however, $\sin^2\Phi$ in the numerator of equation (\ref{result}) tends to zero as the angle approaches either 0 or $\pi/2$, hence the maximum generation efficiency for the paramagnetic to ferromagnetic transition is obtained near $\Phi = \arcsin\frac{1}{\sqrt{3}} \approx 35^{\circ}$. \cite{Buchelnikov92}

EMAT efficiency in single crystal hcp ferromagnets exhibiting helical and fan phases, with external magnetic field applied in the basal plane, has been compared for theoretical and experimental results on Dy.\cite{Buchelnikov94} For an hcp crystal with an ultrasonic wave propagating along the $b$ axis, an external field applied in the basal plane, and an EMAT producing an oscillating magnetic field parallel to the $a$ axis, it was found that in weak external magnetic fields the EMAT efficiency for longitudinal and shear wave generation is given by $\eta \sim M^2$, which is proportional to the magnetic field intensity, and remains small at weak fields. In strong external magnetic fields only a shear wave was generated, with $\eta \sim \chi^2 M^2$; both $\chi$ and $M$ depend on $H$, with $\chi$ decreasing and $M$ increasing with increasing $H$. In the intermediate field regime the authors predicted that the efficiency should first peak, and then decrease as the field is further increased into the strong field regime\cite{Buchelnikov94}, with a strong peak in EMAT efficiency around the helical-fan phase transition due to  domain wall displacements between the fan and helical phase regions.

\section{Experimental details}

The experimental setup and automated data collection and handling has been described in detail in \cite{RachelConf}, and is summarised here. The ultrasonic measurements were done in a pulse-echo configuration \cite{Luthi05b,RachelConf} using a MATEC 6600 pulse generator-receiver, a commercially available X-cut quartz transducer and a custom-designed EMAT.
The EMAT coil was of the single layer spiral (pancake) type, 4.5 mm in diameter, wound using 0.14 mm diameter enamel-coated copper wire. Temperature and magnetic field control were provided by an Oxford Instruments cryostat and superconducting magnet system, with the magnetic field applied to the sample acting as a bias magnetic field for the EMAT.  Sample temperature was measured using a Cernox probe in thermal contact with the sample.  Despite relatively high peak output power, EMATs can be used for low temperature measurements without causing significant problems with heating, because of their low duty cycle \cite {RachelConf,Dobbs73b}. The base temperature during experiments did not differ, within errors, for generation by quartz or EMAT.

A single crystal of  Gd$_{64}$Sc$_{36}$ was cut to a cuboid of dimensions 4.44 x 5.06 x 4.21 mm$^3$ with faces parallel to the main crystallographic directions. An X-cut quartz transducer and an EMAT were placed on opposite sides of the sample, for longitudinal wave propagation along the crystal $c$ axis, with the quartz transducer bonded to the sample using IMI 7031 varnish which also served as a coupling layer. The EMAT was placed in close proximity to the sample, and the change in liftoff between EMAT and sample was negligible during the measurements.

Longitudinal waves at a frequency of approximately 15~MHz were propagated along the $c$ axis of the crystal, probing the $C_{33}$ elastic constant \cite{Luthi05b,Silva99}. This acoustic mode changes interatomic spacing along the $c$ axis and thus is strongly coupled to the magnetic structure, and is particularly sensitive to magnetic phase transitions \cite{Silva99}. The generation and detection of ultrasound in each measurement was done using either the quartz or EMAT. A low-pass filter and averaging were used in all configurations to improve signal to noise ratio. Data collection and offline analysis were automated using LabVIEW.  Three parameters were monitored at different magnetic fields and temperatures; elastic constant $C_{33}$, attenuation of ultrasound, and the efficiency of ultrasound generation (for EMAT only). The elastic constant was obtained by measuring the time between  echoes using cross-correlation of the first echo with the echo train to improve precision \cite{RachelConf,Pantea05}. Attenuation was obtained by fitting echo amplitudes to an exponential decay curve.

\begin {figure} [h]
 \includegraphics[width=70mm]{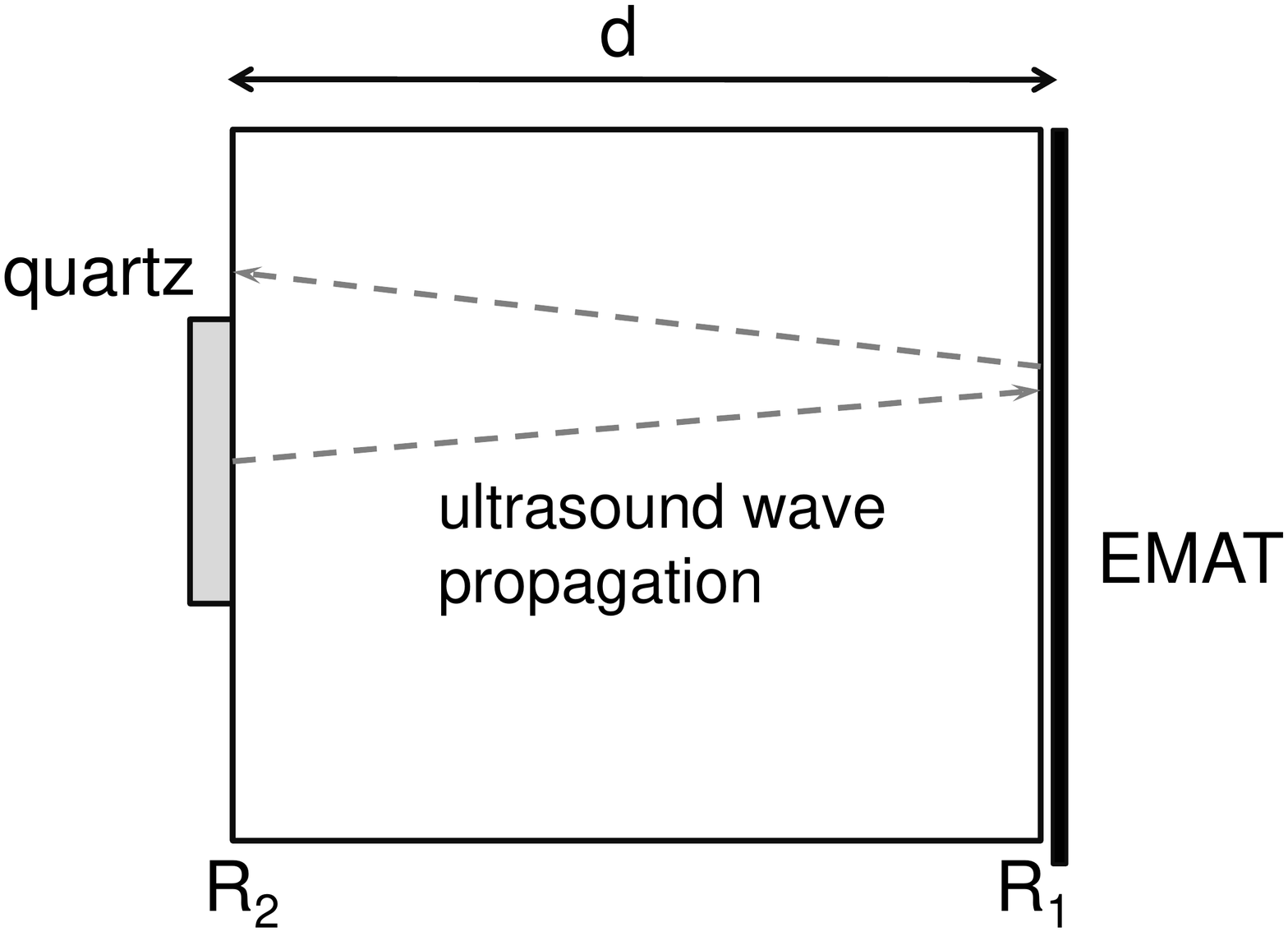}
 \caption { \label {def} Experimental arrangement of quartz and EMAT transducers on the sample. R$_ i$ - reflection coefficients from the corresponding faces of the sample.}
\end {figure}

\section{Attenuation of ultrasound using different transducer-types}\label{Section:efficiency}
Attenuation measurements using contact and non-contact ultrasonic methods can be described mathematically in the following way. Consider a measurement with a quartz transducer and an EMAT on opposite sides of a sample, where $R_1$ is the reflection coefficient for a longitudinal wave from the face near the EMAT, and $R_2$ is the reflection coefficient from the face to which the quartz transducer is coupled (Figure \ref{def}). These coefficients account for scattering and other losses, including losses within the coupling layer for the quartz transducer. As the thickness of the coupling layer is $\ll$ the ultrasound wavelength, only one reflection from the interface with the quartz is considered.  The amplitude of the first echo, $A_1^{(q, e)}$, for quartz or EMAT generation respectively, can be written as
\begin{equation} \label{eampl1}
  \begin{array} {l}
         A_1^{q} =K_q G_q D_q R_1(1-R_2) \  e^{-\alpha 2d}  \\
        A_1^{e} =K_e G_e D_e R_2 \  e^{-\alpha 2 d}
  \end{array}
\end{equation}
where $K_{q, e}$ is the measurement system amplification, $\alpha$ is the absolute attenuation in the sample, $d$ is the sample thickness, $G_{q,e}$ and $D_{q,e}$ are the generation  and detection efficiency, and $(1-R_2)$ is the approximate transmission through the coupling layer. The amplitude of the second echo is:
\begin{equation} \label{eampl2}
  \begin{array} {l}
          A_2^{q} \quad= K_q G_{q} D_q R_1^2 R_2 (1-R_2)\ e^{-\alpha4d}   = A_1^{q}R_1 R_2 \  e^{-\alpha 2d}   \\
         A_2^{e} \quad= K_e G_{e} D_e R_1 R_2^2\ e^{-\alpha4d}   = A_1^{e}R_1 R_2 \  e^{-\alpha 2d}.
  \end{array}
\end{equation}
By fitting experimental echo amplitudes to an exponential decay, the experimental attenuation $\alpha'_{q,e}$  is obtained
\begin{equation} \label{eattn1}
      A_2^{q,e} = A_1^{q,e}\  e^{-\alpha'_{q,e} 2d}.
\end{equation}
This measured attenuation relates to the absolute attenuation via
\begin{equation} \label{eattn2}
 \alpha = \alpha'_{q,e} - \frac{1}{2d}ln(\frac{1}{R_1R_2}).
\end{equation}

For the case of a fully non-contact measurement, the reflection coefficients can be calculated from the acoustic impedances of Gd$_{64}$Sc$_{36}$ and air, and are $R_1 = R_2 = 0.99989$, hence the second term in eq. (\ref{eattn2}) is  $\ll \alpha'_e$, and the measured attenuation in the sample using solely an EMAT is close to the absolute attenuation. When using contact methods the reflection losses are non-negligible; in the case of a quartz transducer coupled to the sample with IMI 7031 varnish, the reflection coefficient $R_2$ may be as low as 0.7, and the correction factor in (\ref{eattn2}) can be of the same magnitude or higher than $\alpha'_q$.  Moreover, the coupling, which is related to the acoustic impedances between the sample, the coupling agent and the quartz transducer, changes as the speed of sound changes \cite{Luthi05b}.  Hence, non-contact EMAT measurements are essential when absolute attenuation is required.

A further difference between measured attenuations (with effects shown in Fig. \ref{behaviour100K}(b)\&(e)) exists due to the relative size of the quartz transducer (diameter 2 mm) and the EMAT (diameter $\approx$ 4 mm). If the acoustic beam  diverges, reflections will not be fully collected by the quartz transducer, resulting in a larger attenuation being measured by the quartz transducer (Figure \ref{def}).  The directivity of a source with diameter $a$ (the largest angular beamwidth) is  $ \sin\theta = 0.61 \lambda/a $.\cite{Kinsler62b}  For a 2 mm diameter transducer operating at 15 MHz this coresponds to an angle of $4^{\circ}$.  The diameter of the beam at the transducer for the first echo expands to 3.17 mm, and 4.35 mm for the second echo, hence there is a difference in the diameters of about 1.4.  Assuming an equal energy distribution across the whole beam, the energy density at the transducer for the detection of the first echo is thus 1.96 times lower, and the detected amplitude 1.4 times lower than would be detected by a large area transducer, and
  \begin{eqnarray} \label{eattn1}
          A_2'^{q} \quad&=& \frac{1}{1.4}A_1^{q}R_1 R_2 \  e^{-\alpha 2d},   \nonumber \\
          \alpha &=& \alpha'_q - \frac{1}{2d}ln(\frac{1.4}{R_1R_2}).
   \end{eqnarray}

Following these considerations, a difference between the attenuation measured by EMAT and quartz transducers of the order of 40 $Np/m$ is estimated. Larger differences may be due to poor electrical contact to quartz transducer and associated losses.

The efficiency of electromagnetic generation of ultrasound can be defined in various ways (see, for example, reference\cite {Dobbs73b}); here we define EMAT  efficiency as the initial generated ultrasonic amplitude, calculated using the amplitude of the first analysed echo and the attenuation.  EMAT detection is similar, but not identical, to generation; in ultrasonic generation the accelerated electrons are moving much heavier ions, while in detection the ions are moving electrons \cite{Palmer03}, and hence the detection is significantly more efficient. For an efficiency difference governed by a factor of $\beta$, from equations (\ref{eampl1})  and (\ref{eattn2}) one obtains
\begin{equation}
G_e = \sqrt{\frac{1}{\beta K_e R_2} A_1^{e} e^{2\alpha d}} = \sqrt{\frac{R_1}{\beta}}\sqrt{\frac{ A_1^{e} e^{2\alpha'_ed}}{K_e}}.
\end{equation}
The system amplification $K_e$ is weakly dependent on EMAT impedance, which may change at a phase transition. This effect cannot be decoupled from the change in EMAT efficiency, but is very weak compared to the overall efficiency and should not impede identification of a phase transition.
Importantly, the unknown coupling to the quartz and the associated reflection $R_2$ is not present in this equation, while  $R_1$, representing reflection from the face near the EMAT, changes negligibly over the phase transition. Hence the first multiplier can be approximated as a constant, and  $\eta_e = \sqrt{ A_{1 \rm{exp}}^{e} e^{2\alpha'_ed}}$, where the experimental value of the amplitude of the first echo includes amplification by the system, $A_{1 \rm{exp}}^{e} = A_{1}^{e} /K_e$.


\section{Results and discussion}

\begin {figure} [h]
 \includegraphics[width=70mm]{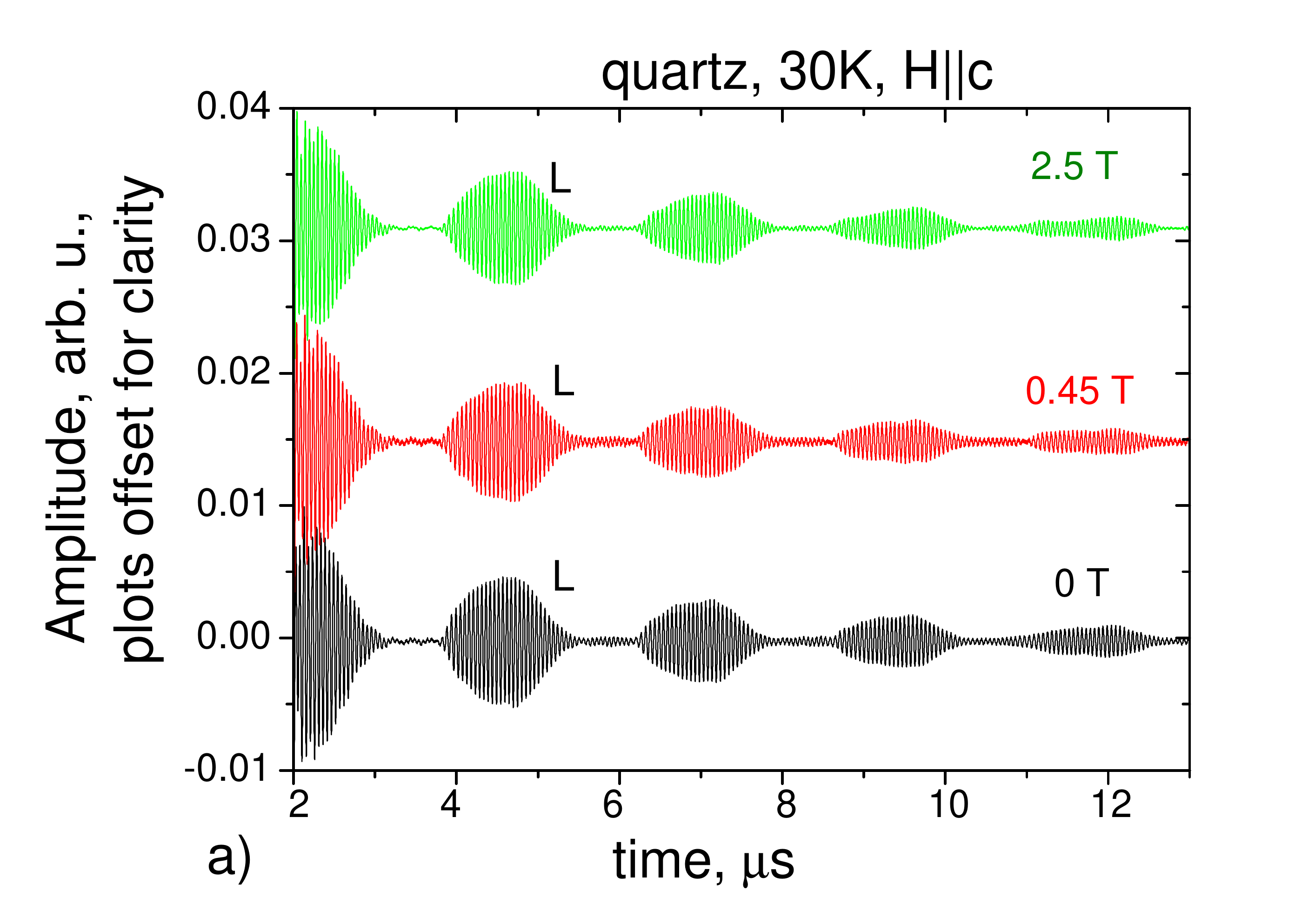}
 \includegraphics[width=70mm]{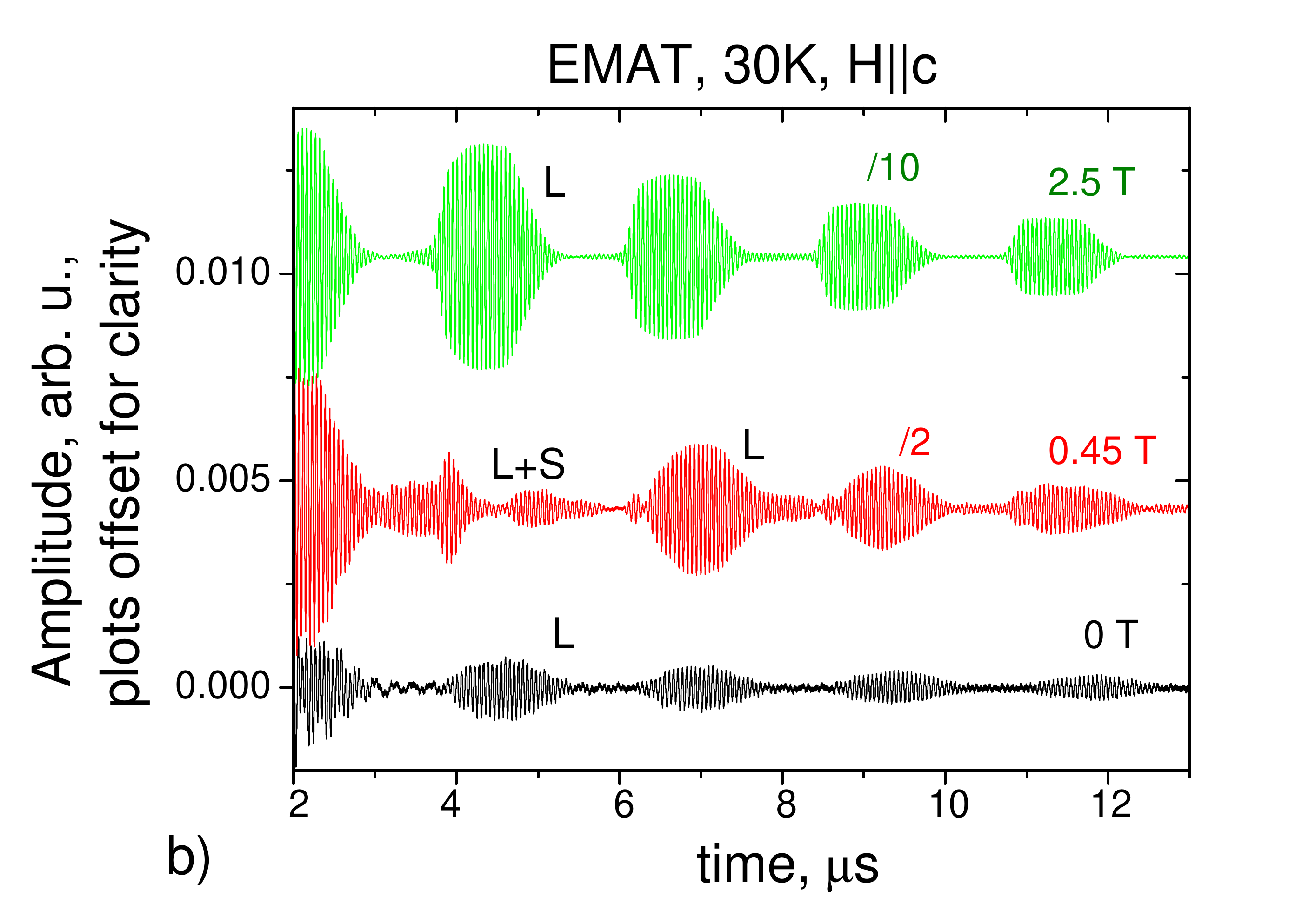}
 \caption {Ultrasound waveforms: echoes propagating along the $c$-axis for  a) quartz transducer and b) EMAT generation and detection,  $\vec{H}\| c$ and $T = 30$~K. The amplitude of the EMAT signals has been scaled as indicated above the traces. \label {echoes}}
\end {figure}

In  experimental configurations with $\vec{H}||c$ and $\vec{H}\bot c$, predominantly  longitudinal-wave EMAT generation was observed, with some  generation and detection at zero applied magnetic bias field (see Fig. \ref{echoes}), characteristic of generation via the magnetostrictive mechanism. Longitudinal waves will be generated in electrically conducting samples via the self-field mechanism, where a Lorentz force is generated due to the time-varying magnetic field from the current pulse, irrespective of a bias field \cite{Jian04}. The self-field is always perpendicular to the sample surface and the coil, and it can be sufficiently large to generate ultrasound in some materials. This effect will also generate ultrasound through magnetostriction for zero bias field. The detection of ultrasound for zero bias field is possible only through the magnetostriction mechanism \cite{Buchelnikov94}.

When a bias field in the direction $\vec {H}|| c$ is applied, the Lorentz force for this EMAT configuration should generate a shear wave with an amplitude linearly dependent on the field magnitude. No such behaviour was observed due to the sample's relatively high resistivity (1040 n$\Omega$m, about 37 times higher than that of aluminium). Hence EMAT generation and detection was predominantly through the magnetostrictive mechanism.

Figure \ref{echoes} shows typical echoes (signal amplitude vs time) obtained using the (a) quartz or (b) EMAT transducer, for different values of the applied magnetic field ($\vec{H}\| c$) at 30~K. For measurements using the quartz transducer there are only small changes in the waveform shape as the magnetic field is increased, and the amplitude of the signal is similar for all traces (allowing for different attenuation). For EMAT measurements the amplitude of the signal changes dramatically between different phases, and in some magnetic phases the EMATs can generate and detect longitudinal and shear waves at the same time \cite{Buchelnikov94}.  As magnetic field is increased, interference of the second and fourth echo of the longitudinal wave occurs due to a wave propagating at about half the speed of the main mode, corresponding to a shear wave. At 0.45 T this interference is particularly strong (middle trace in Figure \ref{echoes}b), with the second shear echo, interfering with the fourth longitudinal echo, becoming very small, due to either a large attenuation of this wave, as expected at a phase transition, or a large change in the shear wave velocity. Because of the interference, the  attenuation and the EMAT efficiency were calculated by taking into account only echoes not visibly affected by the interference. The elastic constant was calculated with and without the interference to allow assessment of where the maximum interference due to this phase transition occurred.

At higher magnetic fields the shear wave is no longer observed. The maximum amplitude of the shear wave will occur close to the phase transitions, as the spin subsystem becomes more mobile and more susceptible to external perturbations \cite{Buchelnikov92}. In the fan or conical helix phases the generation efficiency for the longitudinal wave grows dramatically, and the shear wave may simply no longer be noticeable. Also, by analogy with ref. \cite{Buchelnikov94}, the shear wave may no longer be generated in the high field regime. This could be confirmed through use of a shear-wave transducer, such as y-cut quartz, for detection with EMAT generation in the high field regime.

\begin{figure*}[htbp]
\includegraphics[width=50mm]{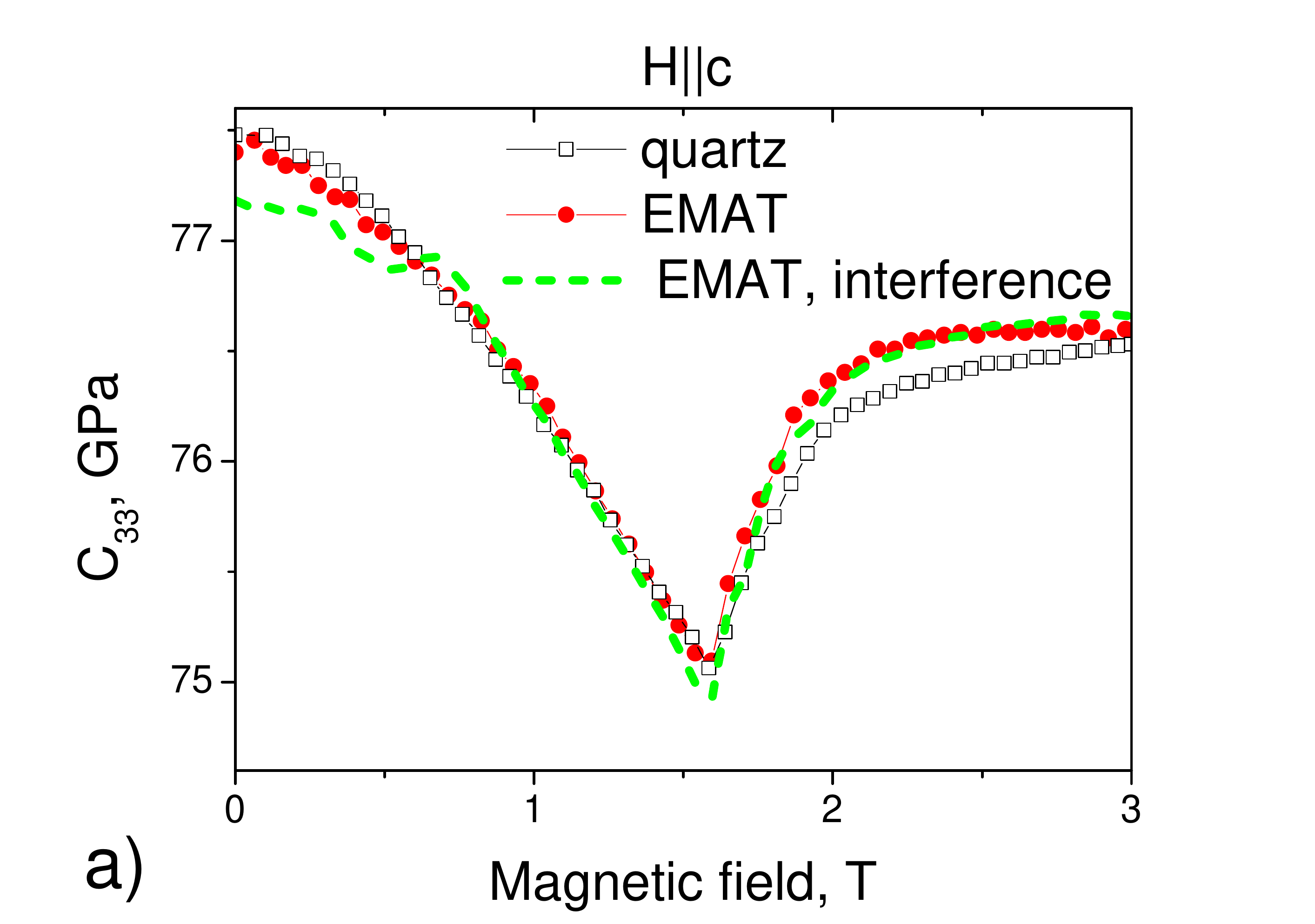}
\includegraphics[width=50mm]{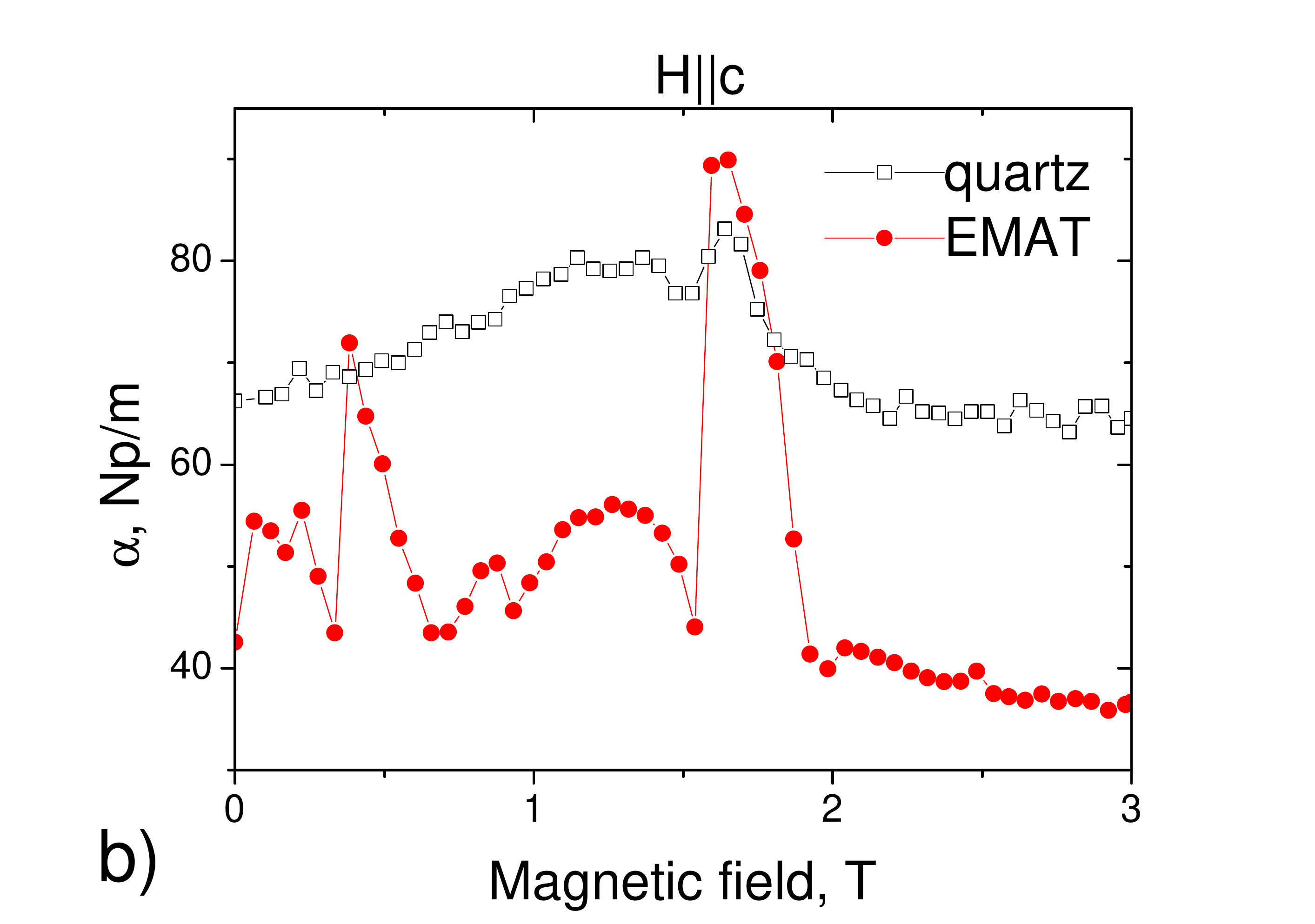}
\includegraphics[width=50mm]{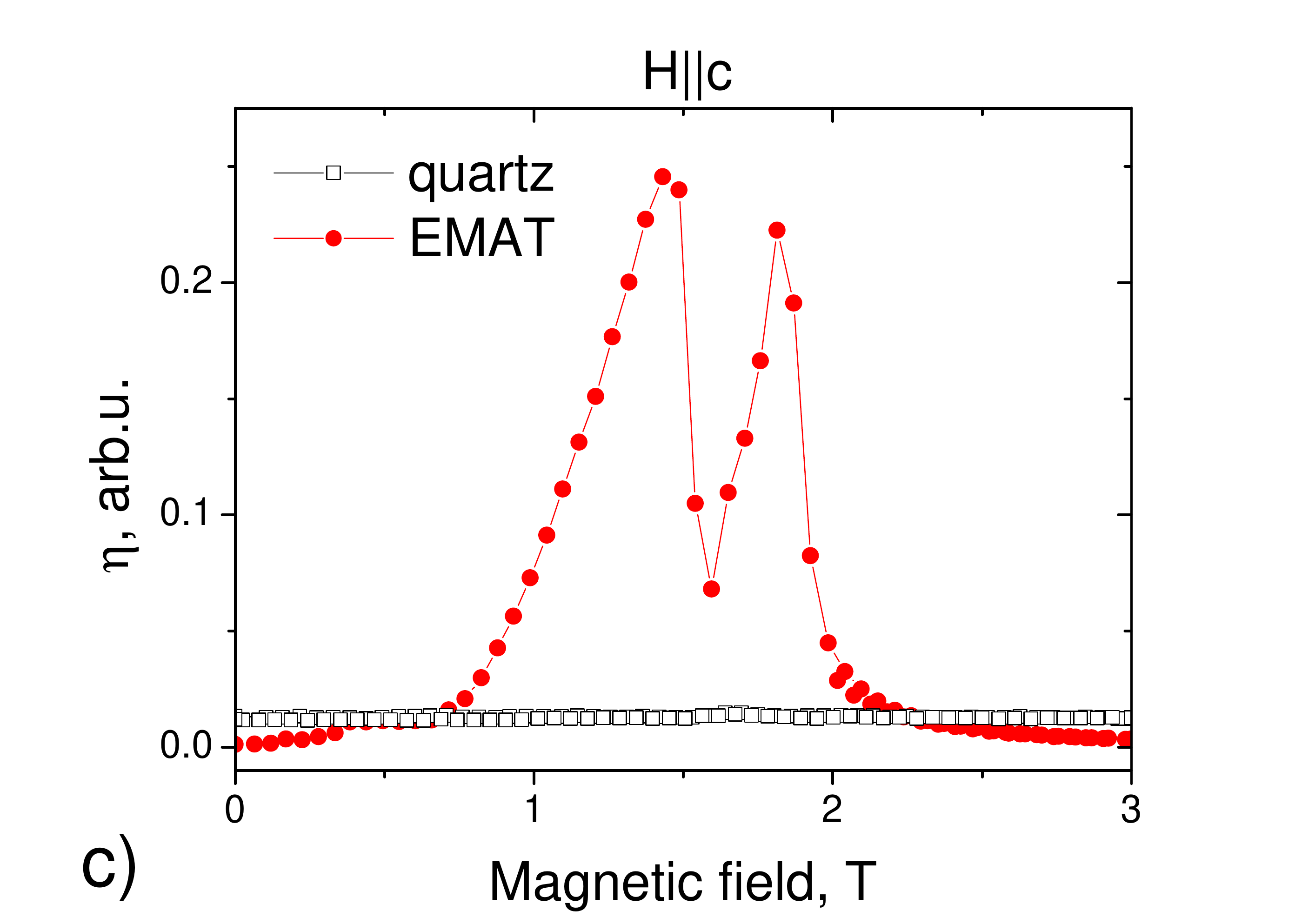}
\includegraphics[width=50mm]{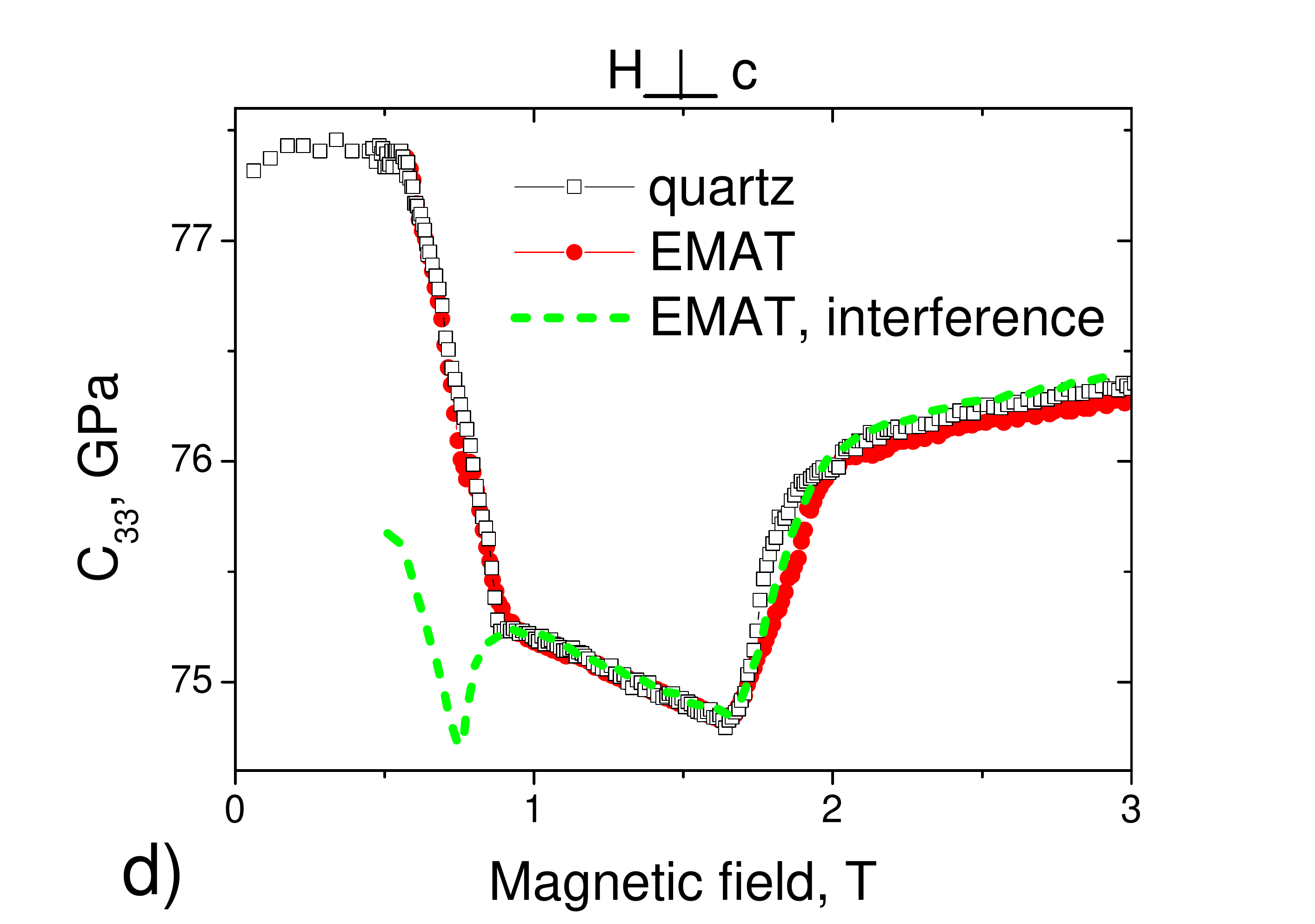}
\includegraphics[width=50mm]{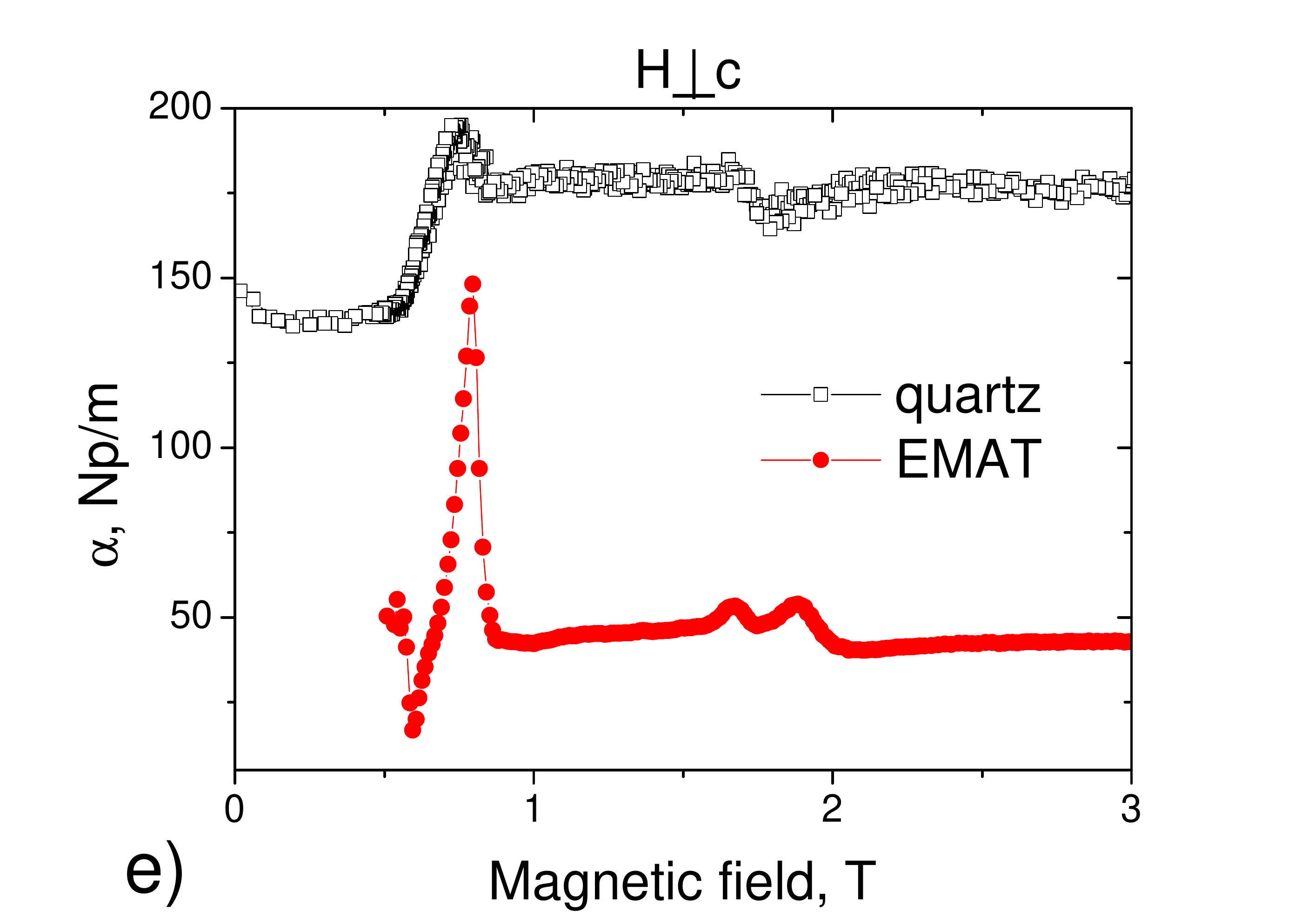}
\includegraphics[width=50mm]{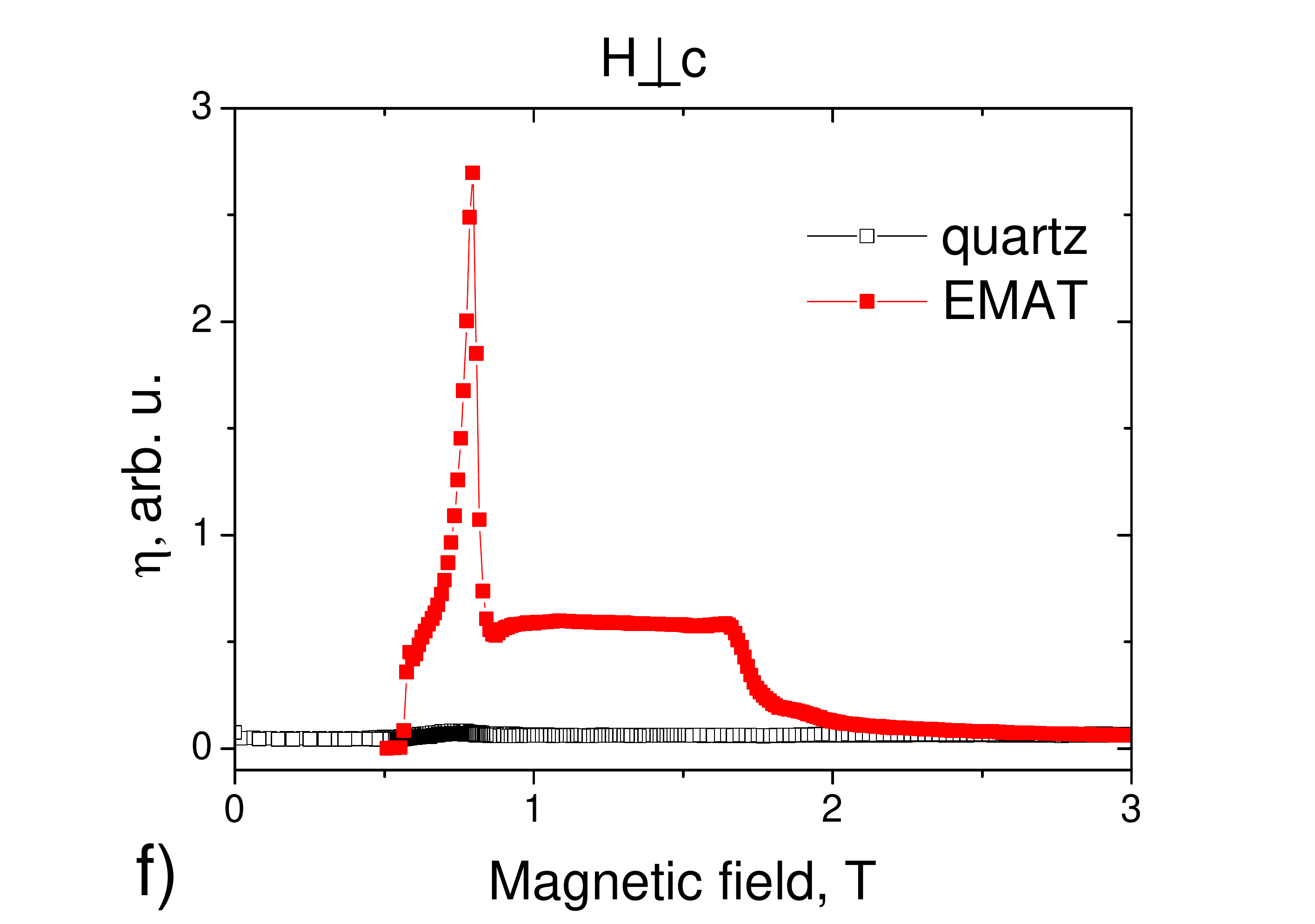}
\caption {Field dependence of elastic constant $C_{33}$ (a, d), attenuation $\alpha$ (b, e) and efficiency $\eta$ (c, f) with $\vec{H}\| c$ (a, b, c) and  $\vec{H}\bot c$ (d, e, f) at 100~K, measured using either the quartz transducer or the EMAT for both generation and detection. \label{behaviour100K}}
\end{figure*}

Figures \ref{behaviour100K}(a)\&(d) show the elastic constant as a function of magnetic field for $\vec{H}\bot c$ (a) and  $\vec{H}\| c$ (d) at 100~K. Open symbols represent measurements using the quartz transducer while closed show EMAT measurements calculated without the interfering echo. As expected, both measurements yield similar results. Two phase transitions can be seen for $\vec{H}\bot c$; at about 0.75 T a first order phase transition between the helix and the fan phases manifests itself as a near-step change in the elastic constant. The field at which the transition occurs can be identified by a maximum in the first derivative of the slope of C$_{33}$, which also coincides with a peak in the attenuation  (Figure \ref{behaviour100K}e) and the EMAT generation efficiency (Figure \ref{behaviour100K}f). At 1.65 T another phase transition, the second order transition between the fan and the field-aligned ferromagnetic phases, can be observed as another significant change in the elastic constant. For $\vec{H}\| c$ (Figure \ref{behaviour100K}a) only one phase transition at about 1.6 T is clear in C$_{33}$, corresponding to the second order phase transition into the field-aligned  ferromagnetic phase.

The dashed lines in Figures \ref{behaviour100K}(a)\&(b) show the elastic constant calculated when including the echo which suffers from the interference. The interference influences the measured time of arrival of echoes, and thus also affects the experimentally obtained elastic constant; this is no longer the true elastic constant, but now shows strongly the position of this lower-field phase transition.  For $\vec{H} \bot c$ the largest error occurs at the phase transition between the helix and fan phases (Figure \ref{behaviour100K}d) \cite{Buchelnikov92}. For $\vec{H} \| c$ the maximum  interference occurs at the same field that the EMAT generation efficiency starts to increase (Figure \ref{behaviour100K}(e)), indicating a phase transition, and the dip in the dashed line in Figure \ref{behaviour100K}(a) in the EMAT measurements was therefore also used as a phase transition point. Note that this transition cannot be seen by a quartz transducer measuring longitudinal waves propagating along the $c$ axis.

The attenuation as measured by the two techniques (Figures \ref{behaviour100K}(b)\&(e)) shows some differences, which can be explained using the discussion in section~\ref{Section:efficiency}.  In the  measurements using the quartz transducer, the attenuation is large compared to that for the EMAT. This is due to the losses associated with the contact measurement, including losses within the couplant, and due to the smaller size of the quartz transducer. For $\vec{H} \bot c$,  the features in the attenuation measured using  the quartz and the EMAT are similar; the first large peak at 0.75~T and a second feature (a step-like change for quartz measurements and the first of the small peaks for EMAT measurements) were chosen as identifiers of the phase transitions based on the expectations from theory \cite{Luthi05b} and with knowledge of the elastic constant. For $\vec{H} \| c$ only one significant peak in attenuation is observed in the measurement using quartz, identifying just one phase transition. In the EMAT measurements the attenuation has an additional feature at 0.45 T.  Although the echo that was visibly affected by the interference was not taken into consideration here, some residual interference with the shear wave may be responsible for this peak. The peak correlated well with the interference-related dip in the elastic constant, and was hence used to determine the phase transition.

As expected, the changes in efficiency of quartz generation and coupling are extremely small and give no useable indication of sample property changes. However, the efficiency of EMAT generation has clear features that coincide with the phase transitions (Figures \ref{behaviour100K}(c)\&(f)). For $\vec{H} \bot c$ the large peak in EMAT efficiency at 0.7 T corresponds to the first order transition from the helix to fan phase. The efficiency stays high in the fan phase and decreases as the sample enters the field-aligned ferromagnetic phase, remaining relatively high. This is similar to the behaviour of  the efficiency in Dy. \cite{Buchelnikov94}

For $\vec{H} \| c$ the efficiency displays very different behaviour. At low magnetic fields the efficiency increases slowly, with a clear onset of significant growth used as the point of entry into the conical helix phase. In this phase the efficiency is relatively high and continues to improve.  The transition to the field-aligned ferromagnetic phase is marked by a dip in the efficiency. This behaviour may be understood as follows; at the transition, the magnetic susceptibility peaks \cite{Buchelnikov92}. The system therefore becomes more susceptible to the external magnetic field, the magnetic moments align strongly along the $c$ axis, and the self-field produced by the EMAT  is no longer able to make a large impact. This could be described by a drop in magnetostriction constant at the phase transition. As the susceptibility decreases past the phase transition point and the alignment becomes less rigid, higher EMAT efficiencies are observed. As the field is further increased and the sample enters deeper into the ferromagnetic phase the efficiency of generation approaches zero, as the magnetic moments become fully aligned along the $c$ axis and the magnetic field produced by the EMAT no longer has any impact on the sample. The dip in the efficiency at the phase transition to the ferromagnetic phase is not observed for $\vec{H} \bot c$ because in this geometry the EMAT efficiently generates ultrasound in the field-aligned phase, so a change in the alignment of magnetic moments due to an increase in susceptibility at a phase transition does not introduce a significant difference to the efficiency.

\begin{figure}[h]
\includegraphics[height=50mm]{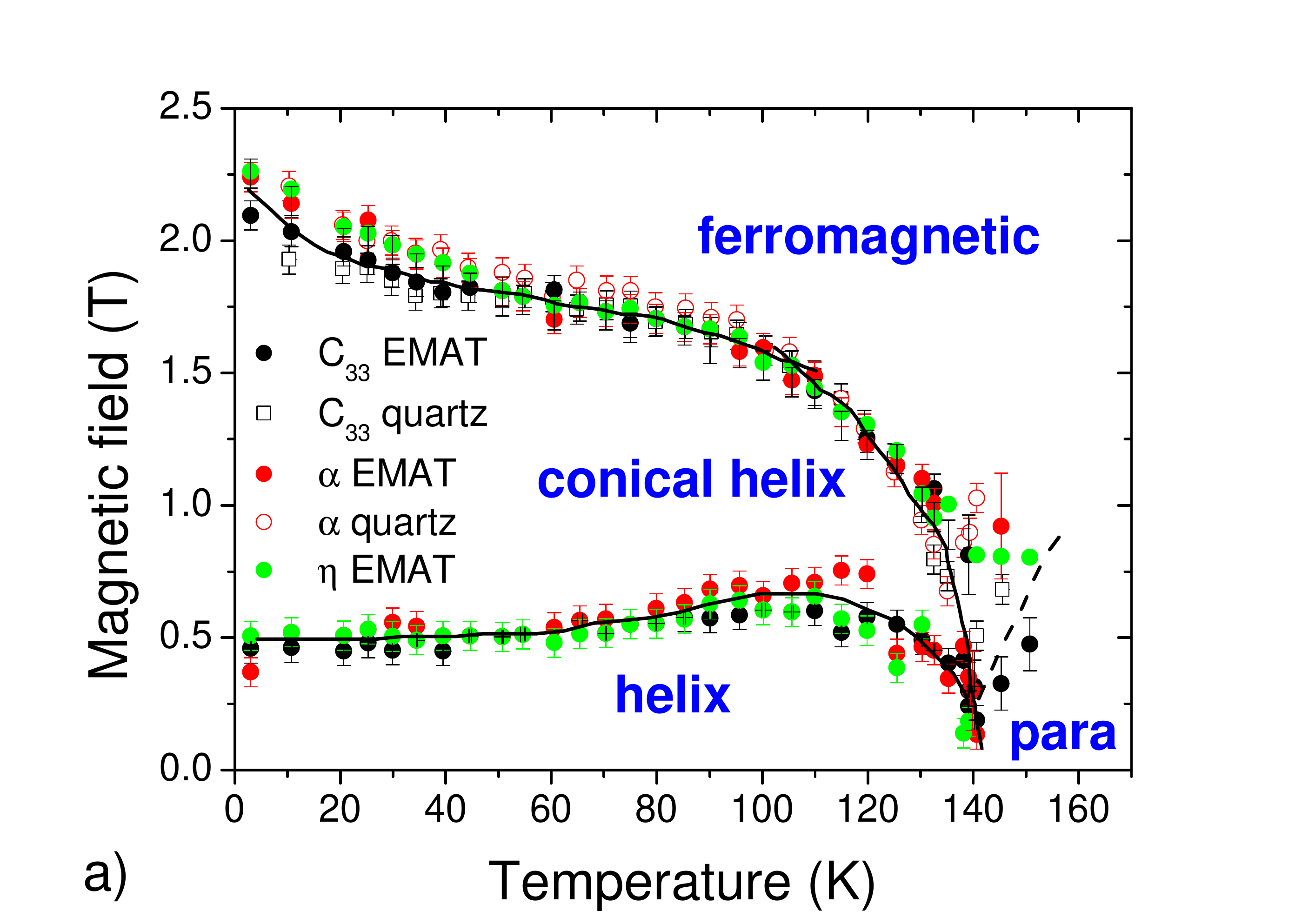}
\includegraphics[height=50mm]{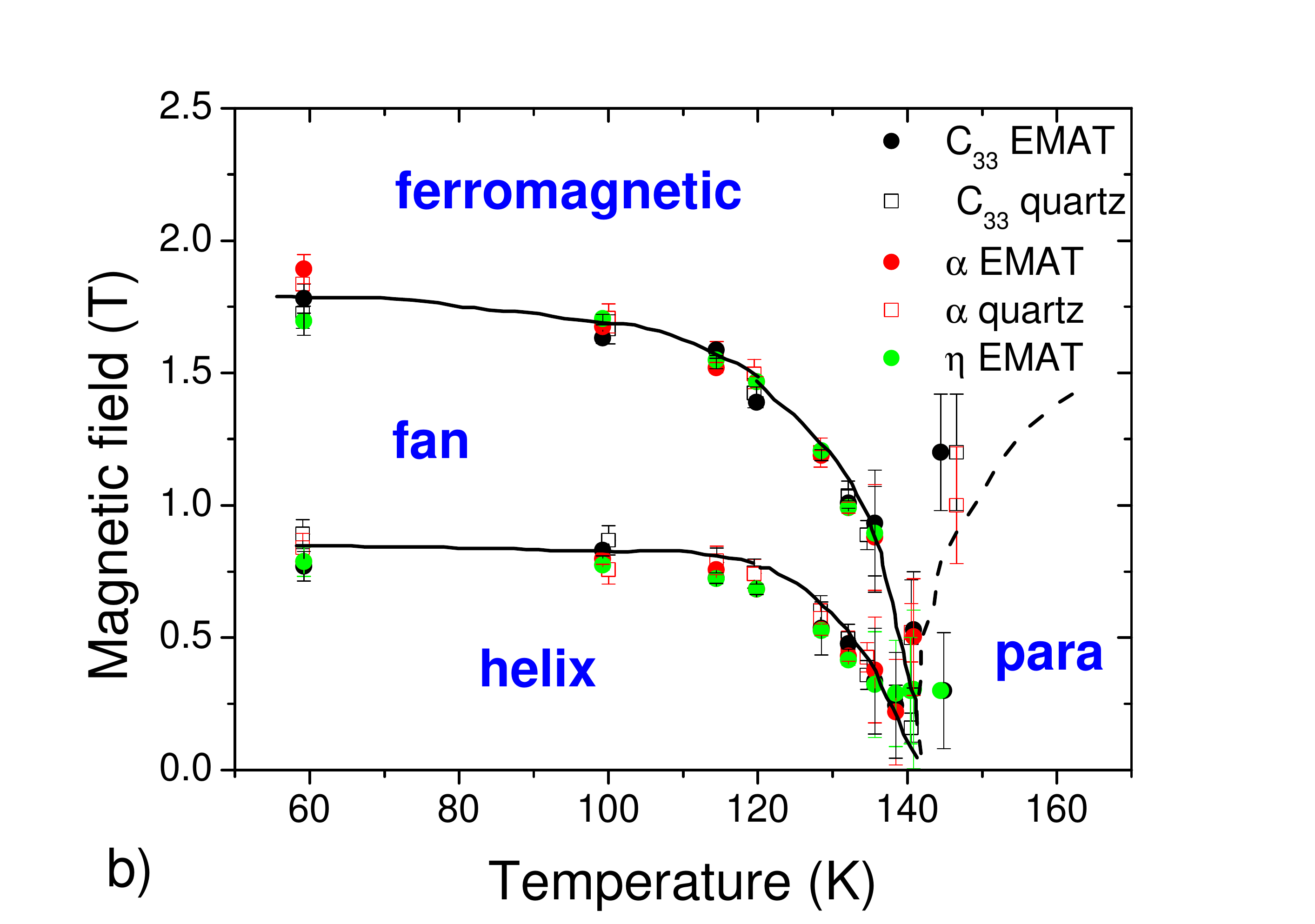}
\caption  { \label{phase} Phase diagram of magnetically induced phases in Gd$_{64}$Sc$_{36}$ with the magnetic field a) $\vec{H} \| c$ and  b) $ \vec{H} \bot c$. }
\end{figure}

The phase diagrams for the two field orientations using all of the above measurements have been mapped and are shown in Figure \ref{phase}. The positions for decreasing magnetic field show a very small hysteresis, but otherwise match those for increasing magnetic field, and have therefore been omitted. Vertical error bars (magnetic field) represent the uncertainty in defining the transition. The error in temperature was calculated individually for each point, and was typically $\pm 0.1K$, smaller than the symbol size on the figure.

The phase diagram for $\vec{H} \bot c$ is consistent with the previous study using contact ultrasonic methods \cite{Silva99}, moreover, we also detect a transition from the paramagnetic to the field-aligned ferromagnetic phase above T$_N$.

\subsection{Electromagnetic generation efficiency}

\begin{figure}[h]
\begin{minipage}[b]{0.5\linewidth} \centering
  \includegraphics [trim = 0mm 0mm 90mm 0mm, clip, height=50mm] {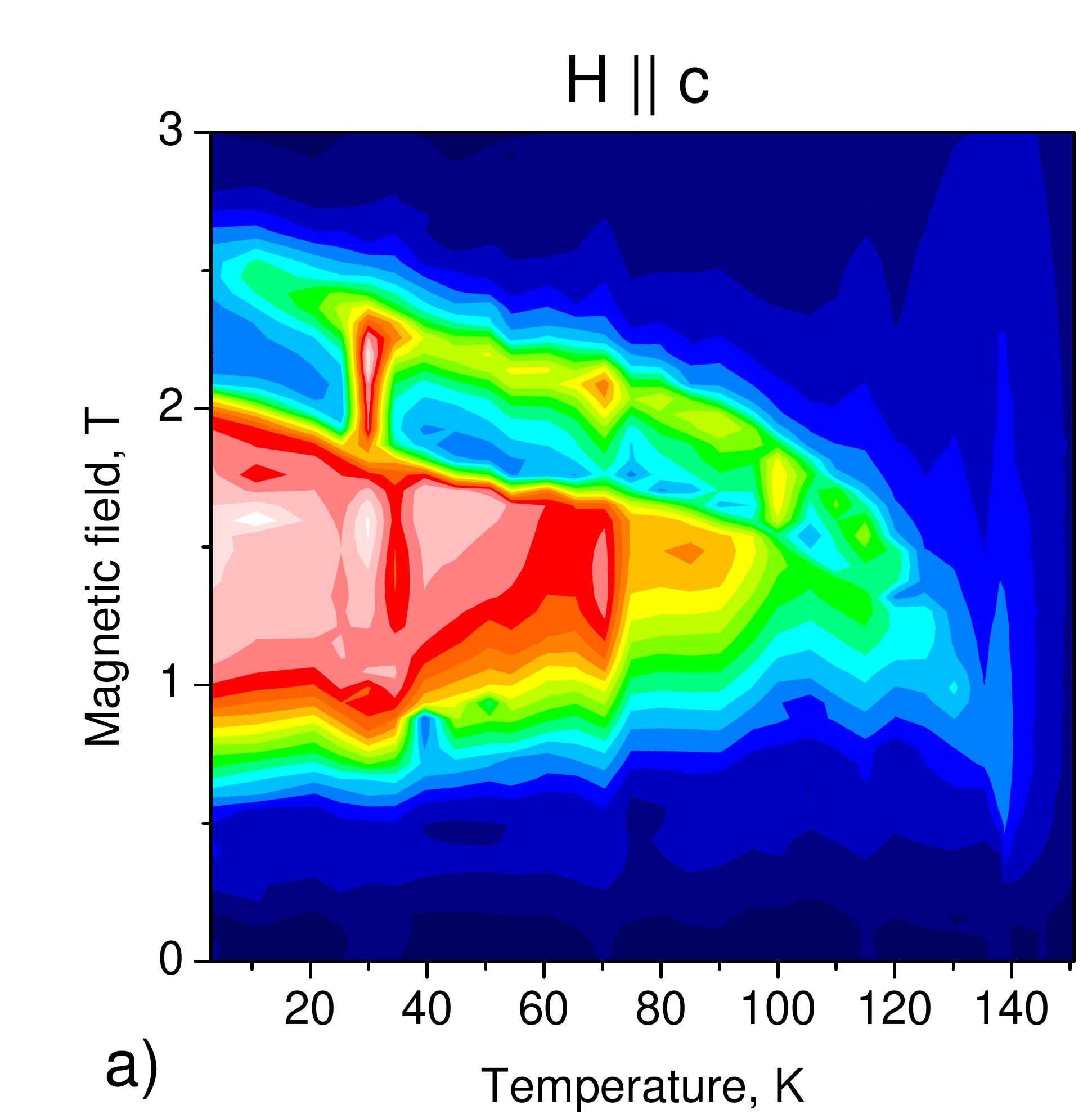}
  \end{minipage} \hspace{0.1cm}
  \begin{minipage}[b]{0.5\linewidth} \centering
\includegraphics [trim = 75mm 0mm 0mm 0mm, clip, height=50mm]{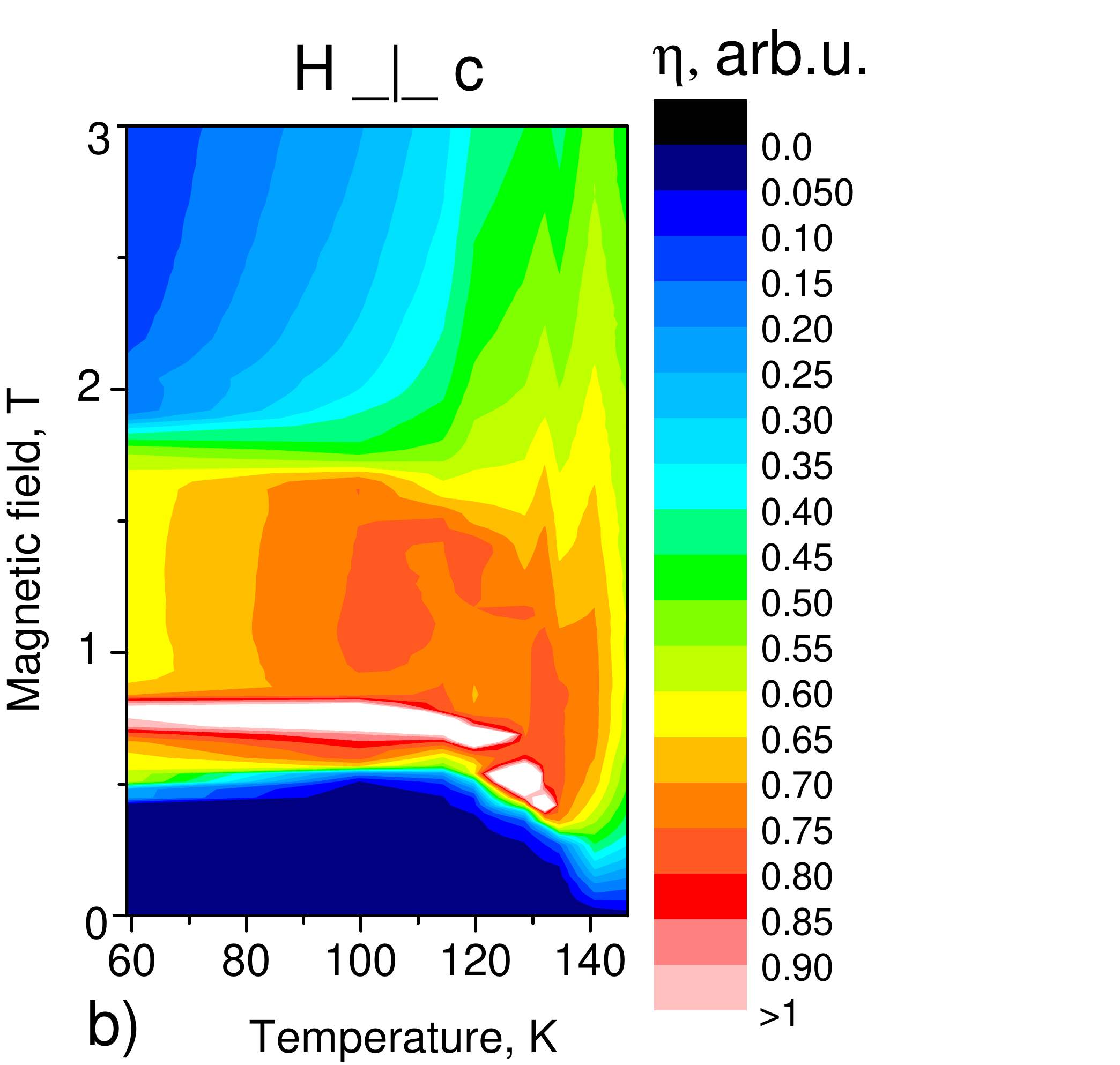}
 \end{minipage}
 \caption {The efficiency $\eta$ of longitudinal ultrasound generation for the two orientations of magnetic field. \label {efficiencymap} The colour scale is the same on both graphs.}
\end{figure}

Figure \ref{efficiencymap} shows the dependence of the efficiency of ultrasound generation on magnetic field and temperature when using the EMAT. Note the resemblance between the maps of the efficiency of ultrasound generation using EMATs (Figure  \ref{efficiencymap}) and the phase diagrams (Figure \ref{phase}). As is apparent from Figure \ref{efficiencymap},  ultrasound generation using an EMAT has a clearly defined region of high efficiency (white, red and yellow area) for each field orientation. For $\vec{H} \bot c$ this high efficiency region corresponds to the fan phase. For $\vec{H} \| c$, the high efficiency region corresponds to the conical helix phase, as discussed above. As the temperature increases to $T_N$, the efficiencies in the fan and conical helix phases decrease while the efficiency in both field-aligned ferromagnetic phases grows. This behaviour of the efficiency for the ferromagnetic phases induced by the applied magnetic field is because magnetostriction tends to zero when magnetic moments make a zero or $\pi$ angle to the longitudinal ultrasound propagation direction  \cite{Buchelnikov92}.   The increase in the efficiency with temperature is due to the increase in thermal fluctuations of magnetic moments, making the the angle between the moments and the generating field deviate from the unfavourable 0 or $\pi$.

\subsection{Demagnetisation and magnetic anisotropy}

Consider a transition to the field-aligned ferromagnetic phase. The magnetic field at which the transition occurs (critical field) indicates how difficult it is to align magnetic moments in the direction of the applied magnetic field, and the value of the critical field will be influenced by the properties of the crystal and by demagnetisation.
The effective field inside the sample is given by $H_{eff} = H_{a} -  H_d$,  where $H_a$ is the applied field, $H_d = NM$ is the demagnetising field, and $N$ is the demagnetising factor.  The demagnetisation factor for a cuboid depends on the magnetic susceptibility as well as the sample shape \cite{Chen05}. Using magnetisation measurements by daSilva et al. \cite{Silva99} for $\mu_0\vec{H}_{\bot c} = 1$ T, the magnetisation in the basal plane  ranges from 2.1 $\mu_B$ per Gd atom in the ferromagnetic phase to 6 $\mu_B$ per Gd atom in the fan phase in the temperature range from 5 to 141 K. Converting this to volume magnetisation following an iteration procedure described in \cite{Chen05}, the corresponding magnetic susceptibility of Gd$_{64}$Sc$_{36}$ ranges from  $\chi = 0.51$ in the ferromagnetic phase just above $T_N$, to $\chi = 1.56$ in the fan phase at 5~K.

\begin{table} [h]
\begin{tabular}{lccccc}
\hline
\rule[-3mm]{0mm}{8mm}
 & $\chi$    &  $N$ & $\mu_0 H_d$ & $\quad N$ & $\mu_0 H_d$\\
\cline{2-6}
\raisebox{2ex}[12pt]{T, K  } &\multicolumn{3}{c} {$\vec{H}\bot c$} & \multicolumn{2}{c}{ $\vec{H}||c$}\\
\hline
\rule[0mm]{0mm}{4mm}

 5      &  1.607   &  0.308  &       0.44 T       & \quad 0.309$^*$  &  0.44 T$^*$ \\
 100  & 1.093   &  0.315  &       0.32 T       &  \quad  0.315$^*$  &  0.32 T$^* $\\
 120  &  0.851   &  0.319 &       0.26 T       &  \quad 0.319$^*$  &  0.26 T$^*$\\
 141 &   0.514   &  0.324  &       0.16 T       &  \quad 0.324$^*$  &  0.16 T$^*$  \\
\hline
\multicolumn{6}{l}{*Assuming similar $\chi$ and $M$ as in $\vec{H}\bot c$ }  \\
\end{tabular}
 \caption {Demagnetisation of Gd$_{64}$Sc$_{36}$ cuboidal sample for $\mu_0\vec{H}_{\bot c} = 1$~T using magnetisation measurements by daSilva et al.\cite{Silva99}.  \label {table}
 }
\end{table}

The sample studied here is a cuboid and has large demagnetisation factors, shown in Table \ref{table} for 1~T applied magnetic field. If susceptibility were the same for directions parallel and perpendicular to the $c$ axis, the demagnetisation factors and fields would be essentially identical for external magnetic fields applied parallel and perpendicular to $c$. However, the crystal is anisotropic and susceptibility and magnetisation are expected to be lower along $c$. Therefore, the demagnetising field  is expected to be smaller for $\vec{H}||c$ compared to for $\vec{H}\bot c$; $H^{\|}_{eff} = H_{a}^{\|} - H_d^{\|}$, $H_{eff}^{\bot} = H_a^{\bot} - H_d^{\bot}$, and $H^{\|}_{eff} > H^{\bot}_{eff}$, with $H_d^{\|} < H_d^{\bot}$.

\begin {figure} [h]
\includegraphics [width=70mm]{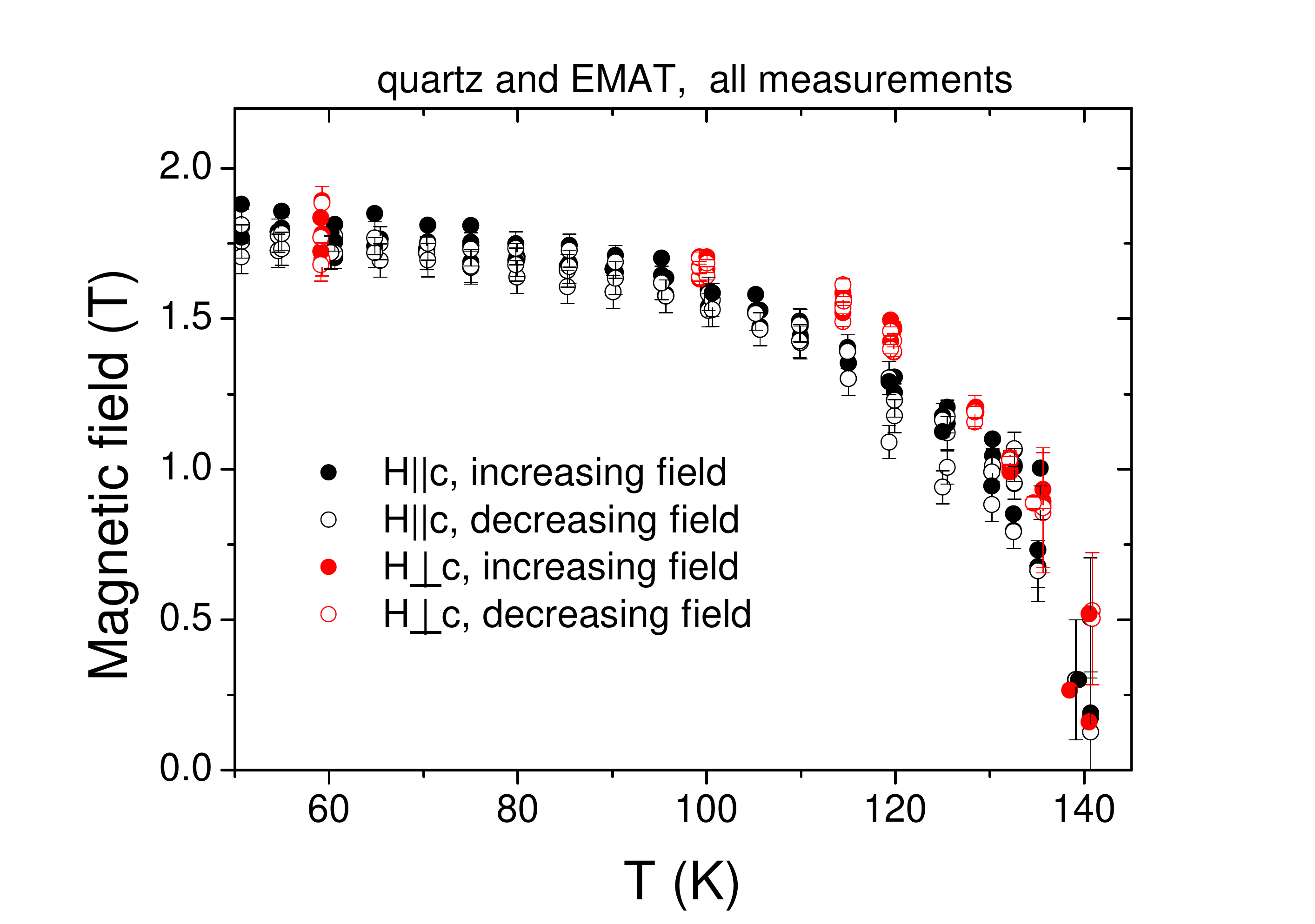}
\caption  {\label {anisotropy} Critical field for the transition to the ferromagnetic phase for magnetic field orientations in the easy magnetisation plane ($\vec{H} \bot c$,  red markers) and perpendicular ($\vec{H} \| c$, black markers), on  increasing and decreasing magnetic field (solid and empty markers correspondingly) as determined by elastic constant, attenuation and EMAT efficiency.}
\end {figure}

One would expect a  critical field $H_{eff}^{\bot}$ (effective field inside the sample) for the transition to field-aligned ferromagnetic phase for $\vec{H}\bot c$ to be lower than $H_{eff}^{\|}$  for $\vec{H}|| c$,  because it is easier to magnetise the crystal perpendicular to the $c$ axis.  Figure \ref{anisotropy} shows that the transition to the ferromagnetic phase for $\vec{H} \bot c$ is at the same field or slightly higher than for $\vec{H} \| c$. These results are consistent for the quartz and EMAT measurements on both increasing and decreasing field. Hence, the effects on the critical fields for the transition to the field-aligned ferromagnetic phase for  $\vec{H} \bot c$ and $\vec{H} \| c$, due to demagnetisation factors and magnetic anisotropy, nearly cancel each other out, and $H_a^{\bot} \approx H_a^{\|}$, with a small difference of 0.15~T.

\section{Conclusions}

Magnetic phase transitions in a  Gd$_{64}$Sc$_{36}$ alloy were studied using two different ultrasonic techniques, the conventional method using contact transducers (quartz), and using electromagnetic techniques (EMATs). Results using quartz and EMAT transducers are consistent with a previous study for $\vec{H}\bot c$ using contact ultrasonic techniques \cite{Silva99}. We suggest a correction to the earlier reported phase diagram, with a second order phase transition from paramagnetic to field-aligned ferromagnetic phase above the Neel temperature (140~K).

The behaviour for $\vec{H}||c$  has been studied for the first time, and for this orientation the intermediate phase between the helix and the field aligned ferromagnetic phases was identified using EMAT measurements, based on attenuation, elastic constants and efficiency maps, and was assigned to a conical helix phase. This phase transition between helix and conical helix was not detected when using a quartz transducer,  which can only probe propagation of one wave-type along the $c$ axis. The EMAT techniques are therefore more sensitive to this phase transition, and potentially to other magnetic  phase transitions, as they can generate and detect longitudinal and shear waves simultaneously, and because the efficiency of EMAT generation through magnetostriction depends strongly on the magnetic state of a material.

EMAT efficiency maps correlate very well with the phase diagrams for the two magnetic field orientations. The efficiency of generation and detection of ultrasound is highest in the fan phase, closely followed by the conical helix phase (Figure \ref{efficiencymap}). This is consistent with the magnetostriction mechanism and magnetisation of the sample and is unique to EMAT generation of ultrasound, and offers insights into the magnetic ordering in the material.

Electromagnetic generation and detection of ultrasound is non-contact, which is particularly important for fragile samples or measurements which require repeated thermal cycling, is suitable for use at cryogenic temperatures,  and potentially allows an absolute attenuation measurement in regions where only one dominant mode is produced. In magnetostrictive samples EMAT generation and detection offers a different way for mapping phase transitions through monitoring the generation efficiency, which is closely related to magnetisation and magnetic phase changes in the materials. This is a fast and simple tool for studying phase transitions in magnetic materials.

\section*{Acknowledgements}
This work was funded by the ERC under grant 202735, NonContactUltrasonic. We would also like to thank A R Clough, D Cleanthous, I J Moore and D J Backhouse for their help with developing and automating the experimental setup. We are grateful to T. Orton for excellent technical support.

\bibliography {ultrasound}

\end{document}